\documentclass[12pt,njp]{iopart}
\usepackage[pdftex]{graphicx}
\usepackage{bm}
\usepackage{epsfig}
\usepackage{dsfont}
\usepackage{bbm}
\usepackage{amsfonts}
\usepackage{amssymb}
\usepackage{amscd}
\usepackage{amstext}
\usepackage{float}
\usepackage{color}
\usepackage{ulem}
\usepackage{enumerate}
\UseRawInputEncoding
\usepackage{iopams}

\usepackage{tikz}

\newcommand{\expec}[1]{\langle #1 \rangle}

\newcommand{\ket}[1]{|\,{#1}\,\rangle}
\newcommand{\braket}[2]{\mbox{$\langle\,{#1}\, | \,{#2}\,\rangle$}}

\newcommand{\sub}[2]{{#1}_{ \mbox{\scriptsize #2}}}

\newcommand{\bv}[1]{\mathbf{ #1 }}

\def\beq{\begin{eqnarray}}
\def\eeq{\end{eqnarray}}

\def\CR{\nonumber\\[0.15cm]}

\newcommand{\rref}[1]{Ref.~\cite{#1}}

\newcommand{\frefp}[2]{figure~\ref{#1}(#2)}

\newcommand{\erefs}[2]{Eq.~(\ref{#1})~(section~\ref{#2})}

\newcommand{\cref}[1]{chapter~\ref{#1}}
\newcommand{\Cref}[1]{Chapter~\ref{#1}}
\newcommand{\aref}[1]{\ref{#1}}
\newcommand{\bref}[1]{(\ref{#1})}

\newcommand\mathplus{+}

\renewcommand{\emph}[1]{\textit{#1}}

\begin{document}

\title{Dynamics of atoms within atoms}
\author{S.~Tiwari$^1$, F.~Engel$^2$, M.~Wagner$^{3,4}$, R.~Schmidt$^{3,4}$, F.~Meinert$^2$ and S.~W\"uster$^1$}
\address{$^1$Department of Physics, Indian Institute of Science Education and Research, Bhopal, Madhya Pradesh 462 066, India}
\address{$^2$Physikalisches Institut and Center for Integrated Quantum Science and Technology, Universit{\"a}t Stuttgart, Pfaffenwaldring 57, 70569 Stuttgart, Germany}
\address{$^3$Max-Planck-Institute of Quantum Optics, Hans-Kopfermann-Stra{\ss}Ÿe 1, 85748 Garching, Germany}
\address{$^4$Munich Center for Quantum Science and Technology (MCQST), Schellingstr. 4, D-80799 M{\"u}nchen, Germany}
\ead{shiva17@iiserb.ac.in, sebastian@iiserb.ac.in}

\begin{abstract}
Recent experiments with Bose-Einstein condensates have entered a regime in which thousands of ground-state condensate atoms fill the Rydberg-electron orbit. After the excitation of a single atom into a highly excited Rydberg state, scattering off the Rydberg electron sets ground-state atoms into motion, such that one can study the quantum-many-body dynamics of atoms moving within the Rydberg atom. Here we study this many-body dynamics using Gross-Pitaevskii and truncated Wigner theory. Our simulations focus in particular on the scenario of multiple sequential Rydberg excitations on the same Rubidium condensate which has become the standard tool to observe quantum impurity dynamics in Rydberg experiments.  We investigate to what extent such experiments can be sensitive to details in the electron-atom interaction potential, such as the rapid radial modulation of the Rydberg molecular potential, or p-wave shape resonance. We demonstrate that both effects are crucial for the initial condensate response within the Rydberg orbit, but become less relevant for the density waves emerging outside the Rydberg excitation region at later times. Finally we explore the local dynamics of condensate heating. We find that it provides only minor corrections to the mean-field dynamics.
\end{abstract}

\maketitle

\section{Introduction}
\label{intro}
The study of quantum impurities has become an important branch of ultra-cold atomic physics, allowing explorations of condensed matter phenomena ranging from the Kondo effect \cite{cyril1997kondo,Nakagawa_kondo_fermionlattice_PhysRevLett}  over polaron formation \cite{grusdt2017strong,camargo2018creation,schmidt2018theory,bruderer2007polaron,bruderer2008transport} to the Anderson orthogonality catastrophe \cite{Knap_imnmpurity_fermions_PhysRevX}. A unique impurity object in this context is a Rydberg atom in a Bose-Einstein Condensate (BEC)  \cite{balewski:elecBEC,gaj:molspecdensshift,Engel_precision_spec,camargo_polarons_exp_PRL,Whalen_Rydbmol_lifetimes_BEC, Mesoscopic:PRL}. Due to the extreme radius of the Rydberg electron density distribution $\sub{r}{orb}\approx 2 a_0 n^2$, which can reach $\sub{r}{orb}\approx 1.8\mu$m at $n=133$, one can enter the realm where tens of thousands of ground-state atoms are located inside the Rydberg orbit and can be set into motion by collisions with the Rydberg electron during the life-time of the latter.

\begin{figure}
\includegraphics[width=0.8\columnwidth]{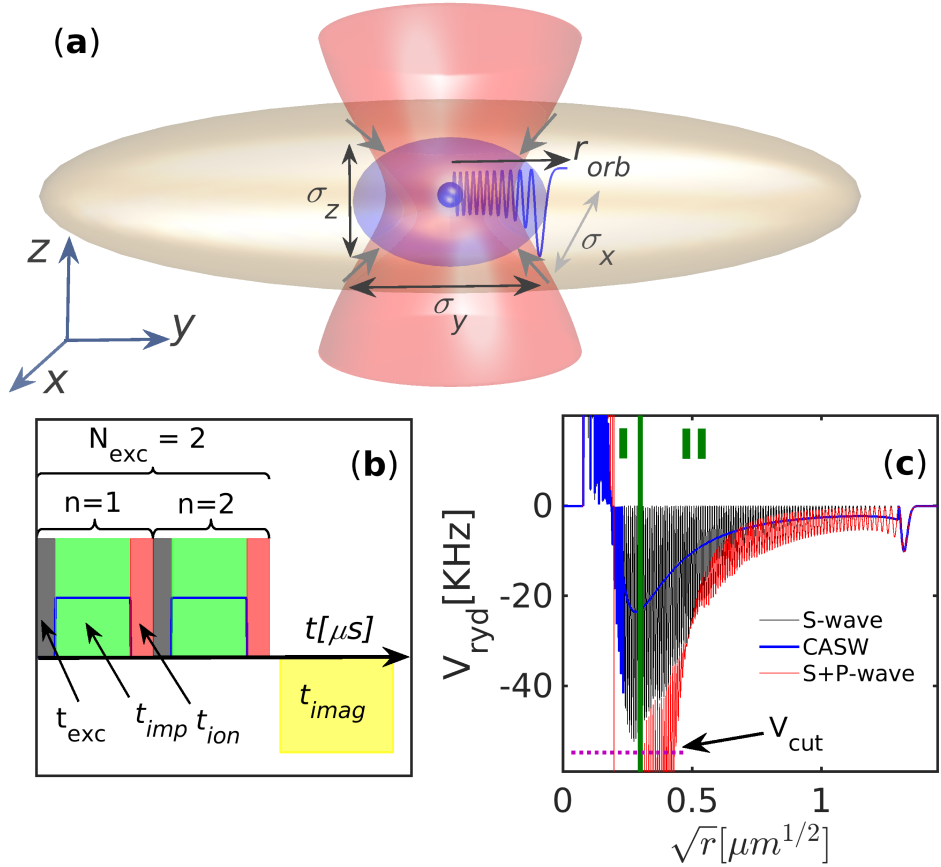}
\centering
\caption{\label{sketch} (a) Sketch of localized excitation of a single Rydberg impurity in the BEC. We assume a Rydberg atom of orbital radius $\sub{r}{orb}$, with its ionic core (dark blue ball) excited in an excitation region (blue shade) within a cigar-shaped BEC (large ellipsoid). The spatial uncertainties of the ion location along the cartesian directions are $\sigma_x$, $\sigma_y$ and $\sigma_z$. (b) Temporal excitation sequence: we assume it takes a time $\sub{t}{exc}$ (grey shade) for the Rydberg atom to be excited, it then remains in the BEC for a time $\sub{t}{imp}$ (green shade) and disappears at the beginning of the field ionization pulse of duration $\sub{t}{ion}$ (red shade). We encode this in the function $\eta(t)$ (blue line), discussed in \sref{electron_atom}, which parametrises presence or absence of a Rydberg atom.
(c) We compare three different levels of detail for the calculation of the interaction potential between Rydberg and ground-state atoms. 
(i) The s-wave interaction potential $\sub{V}{ryd,S}$ (black line) between the condensate and Rydberg impurity, with the nucleus of the latter at $x=0$. (ii) The classical approximation of s-wave (CASW) $\sub{V}{c}$ (blue lines), based on the classical electron probability distribution and (iii) the most complete s+p-wave potential $\sub{V}{ryd,S+P}$ (red lines), which we cut off at $\sub{V}{cut}$ (dotted purple lines). The radial axis is divided into two region (I and II) at $\sub{R}{min}$ with a vertical green line, for the subsequent analysis. The S and S+P potentials agree for $\sqrt{r}\geq 0.7$ $\mu$m$^{1/2}$.
}
\end{figure}
The Rydberg electron affects the Bose-Einstein condensate by imprinting a phase on its mean field wave function at short times, which evolves into density waves at later times. It has been suggested to use these features for tracking the motion, detecting the position and deducing or decohering the quantum state of isolated Rydberg impurities \cite{mukherjee:phaseimp,tiwari:tracking,Karpiuk_imaging_NJP,rammohan:superpos,rammohan:tayloring}. However, for these proposals a detailed understanding of the joint BEC and Rydberg impurity dynamics is required. Experiments probing both to date rely on the repeated excitation of a Rydberg atom at roughly the same location within the laser beam waist \cite{balewski:elecBEC} and then observe the cumulative effect of these excitations. 

In this work we numerically analyze how Rydberg atoms excited within a BEC affect the latter, discussing multiple effects that go beyond earlier studies \cite{Mesoscopic:PRL, Karpiuk_imaging_NJP, Rydberg_Fermi:PRR}. Most importantly we segregate the different condensate dynamics induced by the small, highly-oscillatory radial features of the Rydberg-ground-state potential, from the slower dynamics due to the classically averaged potential on larger scales. We also assess the importance of details of the Rydberg-condensate interaction that only occur very close to the ion core, and thus only affect a relatively small fraction of the interaction volume, such as the p-wave shape resonance. 
Finally, we simulate the local Rydberg induced scattering of atoms from the condensate into the thermal cloud using truncated Wigner theory which allows a separation of condensed and uncondensed components.

Earlier studies have shown that a Rydberg impurity excited in a BEC will result in condensate heating \cite{balewski:elecBEC,Karpiuk_imaging_NJP,tiwari:tracking} that increases with the number of repeated excitations. Since the Rydberg electron interacts only with the condensate atoms present in its orbit, this effect is, on short time-scales, localized within the orbit, and, as we show here, 
not strong enough to invalidate mean-field theory. The small scale oscillations of the radial part of the interaction potential between Rydberg electron and condensate lead to fast condensate dynamics within each radial well, concomitant with a slower, more directed inwards motion governed by the potential envelope. Fast dynamics significantly contributes to the envelope of the density for repeated excitations, which makes it important to accurately model the radial part of the impurity potential. 

Our assessment of the impact of details in the Rydberg-condensate interaction potential is important for future numerical simulations of this problem since those are challenging: the highly oscillatory potential inherited from the Rydberg wavefunction necessitates fine spatial computational grids, while the outwards travelling condensate excitations at later times require a large spatial range.

This article is organized as follows. We introduce our
model of a BEC interacting with a Rydberg impurity via different interaction potentials in \sref{interaction}, and discuss the scheme for the temporal sequence of exciting Rydberg impurities. We then show in \sref{single} that for a single Rydberg impurity, the condensate displays fast dynamics in the small wells of the s-wave impurity potential and a slow average contraction of the condensate particles towards the excitation region. We further compare in \sref{details} the condensate perturbation caused by different interaction potentials and show that the qualitative signal of the condensate is largely insensitive to these details. Out of three potentials used, we justify the use of a cutoff for the full s+p-wave interaction potential in \sref{cutoff}, which has a deep central dip near the ionic core of the Rydberg atom. In \sref{repeated}, we show how the contrast in the condensate response increases through a sequence of multiple Rydberg excitations.       
We go beyond the mean-field approach in \sref{TWA} and show that Rydberg excitations only lead to local heating of the BEC, which, even there, only result minor corrections to mean-field theory. We finally summarize our results in \sref{conclusion} and give an outlook to future directions.

\section{Interactions between Rydberg atom and BEC} 
\label{interaction}
%
We consider a gas of $N$ Rb$^{87}$ atoms with mass $m$ forming a Bose-Einstein condensate. Among the $N$ atoms at most one atom at a time may be excited to a Rydberg state $\ket{\psi} = \ket{\nu s}$, with principal quantum number $\nu$ and angular momentum $l=0$. In this article, we focus exclusively on $\ket{\psi}=\ket{133s}$. The Rydberg excitation shall be quite localized near the origin, as sketched in \frefp{sketch}{a}.  This can be achieved by using a tightly focused excitation laser, additionally exploiting the background density dependent energy shift \cite{balewski:elecBEC}, both of which, however, leave a residual uncertainty $\sigma$ of the impurity position $\bv{x}$, which we explicitly take in account in our model. Similar to experiments \cite{balewski:elecBEC}, we shall also 
consider temporal sequences of single Rydberg excitations with subsequent removal by controlled field ionization.
The excitation, short presence and then removal of the Rydberg atom are repeated $\sub{N}{exc}$ times (see in \fref{sketch}{b}). While the Rydberg atom is present, it remains at the location $\bv{x}_n$, where $n$ is the index of the excitation segment.

\subsection{Electron-atom scattering} 
\label{electron_atom}
%
The interaction of the Rydberg impurities with condensate atoms inside their orbit can be well described by the Fermi pseudopotential \cite{Fermi:Pseudo,greene:ultralongrangemol},
\begin{eqnarray}
\sub{V}{ryd,S}(\bv{R},t) =\eta(t)V_0(\bv{R} - \sub{\bv{x}}{n(t)}))|\psi(\bv{R} - \sub{\bv{x}}{n(t)})|^2.
\label{Swave}
\end{eqnarray}
Here $\eta(t)=1$ at times where a Rydberg impurity is present and $\eta(t)=0$ otherwise, see \frefp{sketch}{b}. 
The shape of the potential is mainly set by the Rydberg electron density $|\psi(\bv{R})|^2$, where $\psi(\bv{R})=\braket{\bv{R}}{\psi}$, since the prefactor $V_0(\bv{R}) = 2 \pi \hbar^2 a_s[k(\bv{R})]/m_e$ only weakly varies as a function of position through the energy dependence of the electron-atom s-wave scattering length $a_s[k(\bv{R})]$. The energy dependent s-wave scattering length is obtained from the phase shift of the electron-atom scattering \cite{Matt:Phase}, with zero energy scattering length $a_s[0] = -16.05a_0$ \cite{Bahrim_2001}. 
For a given Rydberg electronic state, the electron-atom collision energy is given by $\hbar^2 k^2(\bv{R})/2 m_e =  E_{\nu^*} + e^2/(4 \pi \epsilon_0 r)$, where $E_{\nu^*} = - \sub{R}{Ryd}/\nu^{*2}$ is the electronic energy for the effective principal quantum number $\nu^*$ including quantum defects \cite{Gallagher:Qdefect1,Gallagher:Qdefect2}; $r=|\mathbf{R}|$, $\sub{R}{Ryd}$ is the Rydberg constant and $m_e$ is the mass of the electron.
  
The pseudopotential \eref{Swave} takes only $s-$wave collisions between electron and BEC atoms into account, which is a valid approximation for a low energy electron, sufficiently far away from the core of Rydberg atom. 
However, due to a shape resonance for electron-$^{87}$Rb scattering in the $^{3}P^{0}$ channel at $0.02$eV \cite{Fabrikant_1986}, this approximation does not hold all the way to the core. To account for that \eref{sketch} has to be extended by incorporating also higher partial waves \cite{Omont:Pwave}. The calculation of the full interaction potential including p-wave scattering is discussed in \aref{app_Rydmol}. Essentially, we find energy surfaces $\sub{V}{ryd,S+P}(\bv{R})$ from the eigenvalue equation (for the ion at the origin)
\begin{eqnarray}
\sub{\hat{H}}{el}(\bv{R})\ket{\varphi(\bv{R})}&=\sub{V}{ryd,S+P}(\bv{R})\ket{\varphi(\bv{R})},
\label{S_P_wave}
\end{eqnarray}
where $\sub{\hat{H}}{el}(\bv{R})$ is the Hamiltonian for the Rydberg electron including a ground-state atom perturber at location $\bv{R}$, and the state $\ket{\varphi}$ is the one having the largest overlap with $\ket{133s}$ at large $R$, see \rref{Kleinbach:thesis}.

The two potentials, with and without p-wave contribution, are shown in \frefp{sketch}{c}. While not visible in the plotted range of energies, the s+p-wave potential is vastly stronger at shorter $R$, reaching to $V\approx -680$ MHz. For numerical stability, we cut this divergence off when $V=\sub{V}{cut}$. To ensure convergent results, we investigate the dependence on $V=\sub{V}{cut}$. One of the objectives of this article is to explore to which extent the detailed shape of the interaction potential and cutoff affects the BEC response. Hence we will compare simulation results arising from both these potentials and varied cutoffs. 

Also the highly oscillatory character that is inherited from the radial part of the Rydberg wavefunction  
poses numerical challenges. This can be partly alleviated by replacing the Rydberg-electron wavefunction with a classical approximation \cite{Amartin}, as discussed in \cite{tiwari:tracking}. To this end, we  consider a third potential, referred to as ``classical approximation of s-wave'' (CASW), in which we replace \bref{Swave} for ion at the origin by
\begin{equation}
\sub{V}{c}(\bv{R},t) = \eta(t) V_0(\bv{R}) 
\cases{
\rho^{Q}(\bv{R})&  $r<\sub{R}{min}/{2},$\\
\rho^{\mbox{cl}}(\bv{R})& $\sub{R}{min}/{2}<r< \sub{R}{ct},$ \\
\rho^{Q}(\bv{R})&  $r\geq \sub{R}{ct},$}
\label{Classpotfull}
\end{equation}
where $\rho^{Q}(\bv{R})=|\psi(\bv{R})|^2$ and $\rho^{cl}(\bv{R})$ (see \aref{app_Rydmol}) are the quantum and classical electron probability densities respectively, as used in \rref{tiwari:tracking}, $\sub{R}{ct}$ is the outer classical turning point and $\sub{R}{min} \approx 0.1\mu$m or $\approx 1900a_0$. The latter is the distance of the Rydberg nucleus from the shape resonance in electron-$^{87}$Rb scattering for the $\ket{\nu l}=\ket{133s}$ Rydberg state, as indicated in \frefp{sketch}{c}. The approximation \bref{Classpotfull} is sketched in \frefp{sketch}{c} as a blue line.

\subsection{Temporal excitation sequence} 
\label{sequence}

The excitation sequence creating a single Rydberg impurity at a time, subsequently removing it by field ionization, and then repeating the cycle is sketched in \frefp{sketch}{b}. Here $\sub{t}{exc}$, $\sub{t}{imp}$, and $\sub{t}{ion}$ are the times taken to excite the Rydberg state, free imprint time, during which the Rydberg atom resides in the BEC, and the time for ionization of the Rydberg impurity, respectively. 
Therefore, the time $\tau$ taken for $\sub{N}{exc}$ repeated excitations to complete before the free evolution time ($\sub{t}{imag}$) of the BEC is $\tau =  \sub{N}{exc}(\sub{t}{exc} + \sub{t}{imp} + \sub{t}{ion})$.

In our simulations for $\sub{N}{exc}>1$, the location of the ionic core of the n'th Rydberg atom $\bv{x}_{n(t)}$ is randomly drawn from a three-dimensional (3D) Gaussian distribution, with Cartesian standard deviations $\sigma_{x,y,z}$. For this we determined parameters  $\sigma_{x,y,z}$
 corresponding to a given excitation laser beam profile and background density profile in \aref{app_Rydexcitation}. Only for simulations with $\sub{N}{exc}=1$, the Rydberg location is at the origin. To model the sequence shown in \fref{sketch}, $\bv{x}_n(t)$ is thus a step-wise continuous vector function, 
assembled from the $\sub{N}{exc}$ random 3D positions $\mathbf{x}_n$ and the integer index $n(t)=\lfloor t/\tau \rfloor$. Finally, $\eta(t)$ is given by $\eta(t)=\sum_{n=0}^{\sub{N}{exc}-1}\bar{\eta}(t - n\tau)$, with
$\bar{\eta}(t)=\left[\tanh(\frac{t-\sub{t}{exc}}{\xi}) + \tanh(\frac{\sub{t}{imp}-t}{\xi})\right]/2$, with near instantaneous risetime $\xi$.

\subsection{Condensate response} 
\label{GPE}
%
In order to understand how the excitation sequence of Rydberg impurities affects the BEC, we model the latter using the Gross-Pitaevskii equation (GPE) and assume an immobile Rydberg atom, thus treating the impurity potential \bref{Swave} as external potential \cite{middelkamp:rydinBEC,balewski:elecBEC,mukherjee:phaseimp,Karpiuk_imaging_NJP,Shukla_pandit_particlesinsuperfluids}.
The resulting GPE including the interaction with the Rydberg impurity is
\begin{eqnarray}
\label{impurityGPE}
&i\hbar \frac{\partial}{\partial t}\phi(\bv{R})= \bigg(-\frac{\hbar^2}{2 m}\boldsymbol{\nabla}^2 + W(\bv{R}) + U_0 |\phi(\bv{R})|^2 \CR
&+\sub{V}{ryd,S,n}(\bv{R},t) + i\hbar \frac{K_3}{2} |\phi(\bv{R})|^4 \bigg)\phi(\bv{R}),
\end{eqnarray}
where $\phi(\bv{R})$ is the condensate wave-function, and $W(\bv{R}) = m(\omega^2_x x^2 + \omega^2_y y^2 + \omega^2_z z^2)/2$ is the 3D harmonic trap, using $\bv{R} = [x,y,z]^T$. The strength of interactions among ground-state $^{87}$Rb atoms in the condensate is set by $U_0 = 4 \pi \hbar^2 a_b/m$, where $a_b = 109 a_0$ is the s-wave scattering length. 
The last term on the right hand side of \eref{impurityGPE} phenomenologically incorporates short-range three-body loss of the BEC. We take a rate constant $K_3 = 1.8 \times 10^{-41} m^6/s$ \cite{Three_body_decay, Three_body_loss:PRL, Three_body_loss:Sebastian, Three_body_loss:Savage}
assuming a hyperfine state $\ket{F, m_F}=\ket{2, 2}$.

The complex-valued condensate wavefunction can be written as $\phi(\bv{R}) = \sqrt{\varrho}e^{i\varphi(\bv{R})}$, with real density $\varrho(\bv{R})$ and real phase $\varphi(\bv{R})$. In the Raman-Nath approximation, the initial effect of the Rydberg impurity is then to imprint a phase $\varphi(\bv{R}) = - V_{ryd}(\bv{R})\Delta t /\hbar$ within a short time $\Delta t$ \cite{dobrek:phaseimp,mukherjee:phaseimp} while leaving the density relatively unaffected. Only with some delay will phase gradients, corresponding to initially imparted momentum, be converted into variations of condensate density through motion of the ground-state atoms \cite{tiwari:tracking}.  
For a numerically and conceptually tractable model, we ignore the backaction on the Rydberg impurity in the present study.

\section{Condensate response in the mean field}
\label{GPE_Single}

\subsection{Single spherical excitation}
\label{single}
%
We first consider the case of a single impurity excited to the $\ket{\nu l} = \ket{133s}$ Rydberg state at $[x,y,z]^T = [0,0,0]^T$ in a homogeneous BEC of density $\rho_0 = 4.0\times10^{14}$cm$^{-3}$. All our results also in later sections pertain to this density. Due to the spherical symmetry of the Rydberg s-state, the entire problem is spherically symmetric, and we can solve the radial GPE instead of \bref{impurityGPE}, which using $\phi(\mathbf{R})=u(r)/{r}$ for $r= |\mathbf{R}|$ becomes \cite{wuester:nova}
\begin{eqnarray}
i\hbar \frac{\partial}{\partial t}u(r,t) &= -\frac{\hbar^2}{2 m}\frac{\partial^2u(r,t)}{\partial r^2}  + \frac{U_0}{ r^2} |u(r,t)|^2u(r,t)\CR
& + \sub{V}{ryd,S,n}(r)u(r,t),
\label{radialGPE}
\end{eqnarray}
where $u(r)$ is a radial condensate wavefunction, normalised such that $\int_0^\infty dr |u(r)|^2=N$. We discuss in \aref{app_radial} how to handle the practical implementation in a homogenous system using the Fast-Fourier-Transform for derivatives. Eq.~\bref{radialGPE} is finally solved using the high level computing language XMDS \cite{xmds:docu,xmds:paper}.

\begin{figure}
\includegraphics[width=0.9\columnwidth]{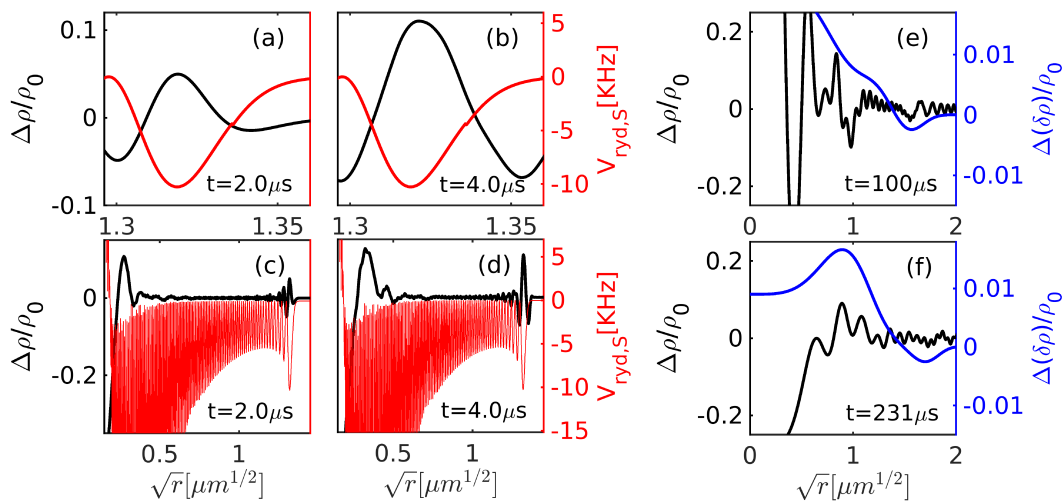}
\centering
\caption{\label{fig_single} Condensate response to a Rydberg impurity in the spherically symmetric case. The left y-axis shows the condensate density relative to the background density $\rho_0 = 400 \mu$m$^{-3}$, i.e. $\Delta \rho/\rho_0 = (\rho - \rho_0)/\rho_0$, and the s-wave interaction potential in Eq.~\eref{Swave} is shown on the right y-axis of panels (a-d). (a) The condensate density (black solid line) at the imprint time $\sub{t}{exc} + \sub{t}{imp} = 2.0\mu$s in the outer radial well (red line). (b) The condensate density in the outer well before ionizing the Rydberg impurity at $\sub{t}{exc} + \sub{t}{imp} = 4.0\mu$s. The density over the whole range of radii is shown in (c) and (d) in the same style as for panels (a) and (b), respectively. The density at late time after some evolution with impurity removed is shown at $\tau + \sub{t}{imag} = 100\mu$s in (e) and $\tau + \sub{t}{imag} = 231\mu$s in (f) as black line.
In (e,f) the solid blue line using the right axis is a radial cut through the 3D spatially smoothened density, using a Gaussian kernel with a standard deviation of $\sigma = 0.5 \mu$m; note the different y-axis scale.
The green arrow in (e) and (f) indicates the inward and outward flow of the condensate, respectively.
}
\end{figure} 
For now, we do not yet consider three-body loss ($K_3=0$) and allow the Rydberg impurity to interact with the condensate through the s-wave interaction potential \bref{Swave}. We show the resulting dynamics of the condensate density deviation from the background density $\Delta \rho=\rho(r)-\rho_0$ in \fref{fig_single} as function of $\sqrt{r}$. The condensate particles respond to each radial well of the s-wave interaction potential initially, accelerating towards its center, shown for the outermost well in the range $1.3 {\mu m}^{1/2} < \sqrt{r} <1.35 {\mu m}^{1/2}$ at $\sub{t}{exc} + \sub{t}{imp} = 2.0 \mu$s in panel (a) of \fref{fig_single}, where $\sub{t}{exc} = 0.5 \mu$s and $\sub{t}{imp} = 1.5 \mu$s. While the relative density contrast near the outer well increased by a factor of about three  within an additional imprint time $\Delta t = 2.0 \mu$s towards panel (b), as long as the Rydberg impurity remains present, the maximal relative density increase reaches only $13$\% in the outer well.

In \frefp{fig_single}{c,d}, we show the density in the full radial range at the same times as in (a,b). While the condensate response to the outer well of the impurity potential is significant, the response to most of the inner wells remains disproportionally small. 
We can understand this from the fact that the healing length here is $\xi = 0.13$ $\mu$m, hence only the outer well is wider than this scale. 
The response to all other wells is suppressed by the inability of the mean field to respond on that length scale.
Altogether, the net effect on the density remains small at the imprint time ($\sub{t}{exc} + \sub{t}{imp} = 4.0 \mu$s), corresponding to the Raman-Nath regime \cite{mueller_raman_nath}. In contrast, on larger time-scales the initial momentum imparted by the Rydberg impurity keeps the condensate flowing towards the origin, leading to a much increased relative density there. We find $\Delta \rho/\rho_0 \approx 2.5$ near the ionic core at $\tau + \sub{t}{imag} = 100 \mu$s for the case of \fref{fig_single}(e), where $\tau = \sub{t}{exc} + \sub{t}{imp} + \sub{t}{ion} = 4.8\mu$s is the time taken to complete one excitation and its subsequent removal ($\sub{t}{ion} = 0.8 \mu$s) as discussed in \sref{sequence}. 

In \fref{fig_single}{e}, we show the spatially smoothened density $\delta\rho$ as a blue line, to demonstrate the net inwards flow towards the Rydberg ion. We obtain the latter as the spatial average of the 3D condensate density over a Gaussian kernel with a standard deviation of $\sigma = 0.5 \mu$m, for this task solving the 3D GPE \bref{impurityGPE} using ${512\times 512\times 512}$ gridpoints and accepting a slightly undersampled Rydberg potential. Through comparison of the 3D simulations with undersampled potential and radial 1D simulations with resolved potential, we ensure that all qualitative conclusions presented here are consistent in both simulations.

After piling up in the centre, the excess condensate density then subsequently diffuses due to atomic inertia and is converted into an outward flow of condensate particles, see \ref{fig_single}(f). Overall we see that the net impact of the excitation is to contract the BEC towards its location, which seems intuitive given the net attractive character of the s-wave potential in \frefp{sketch}{c}, when spatially averaged over all the wells.


\subsection{Dependence on potential details}
\label{details}

There are two main features of the potentials in \frefp{sketch}{c} that can be considered at varying levels of approximation: 
\begin{enumerate}[(i)]
\item The rapid oscillations stemming from the radial Rydberg wavefunction could be removed when replacing the potential with the CASW variant \bref{Classpotfull}, while leaving the net attractive character intact. This replacement can help to keep numerical simulations of larger condensates tractable. Since the characteristic length scale for mean-field condensate dynamics is given by the healing length $\xi=0.13\mu$m, which is much larger than the radial oscillation wavelength of the exact potential, one can except the effect of those oscillations to be somewhat averaged out, as we confirmed in \sref{single}. 
\item While the s-wave and the s+p-wave potentials agree for $r>\sub{R}{min}$, they deviate significantly closer to the nucleus. However, in three dimensions, the corresponding volume is a small fraction of the Rydberg orbital volume ($5\%$ at $n=133$), and hence the importance of this difference for the BEC dynamics at larger distances from the nucleus is not a priory clear.
\end{enumerate}

\begin{figure}
\includegraphics[width=0.8\columnwidth]{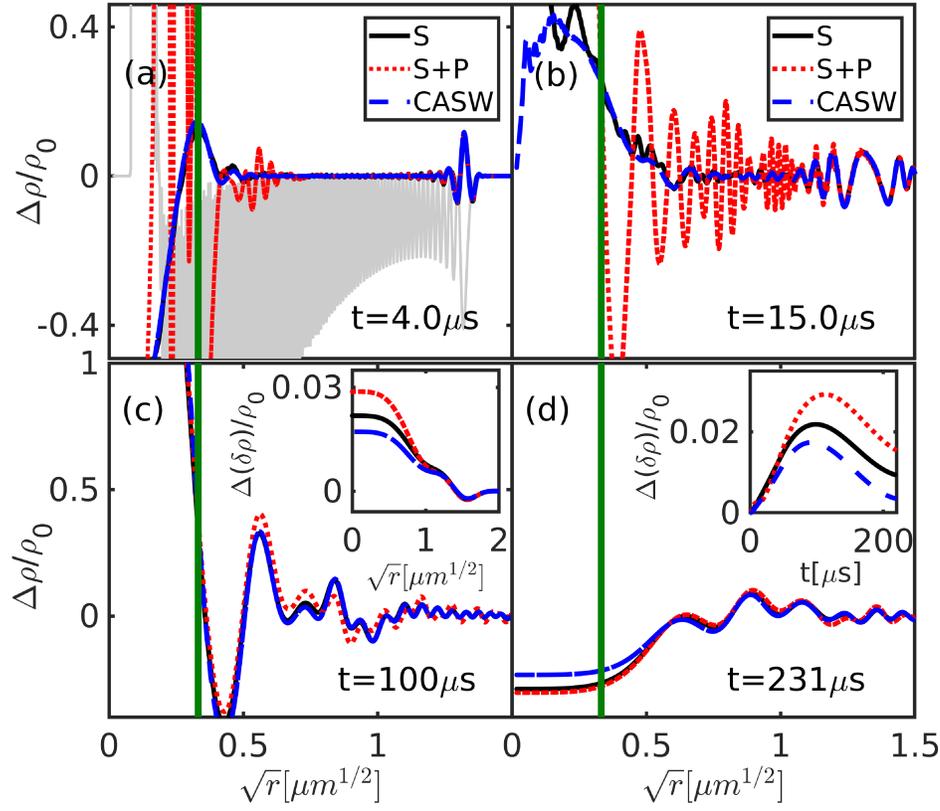}
\centering
\caption{\label{fig_single_comp} Motion of the BEC in different approximations of the Rydberg interaction potentials. The condensate density is shown relative to the background density as in \fref{fig_single}. (a) Atom density in interaction potentials, s-wave (solid black line) Eq.~\ref{Swave}, s+p-wave (dotted red line) Eq.~\ref{S_P_wave}, and CASW (dashed blue line) Eq.~\ref{Classpotfull} are shown at the imprint time $\sub{t}{exc} + \sub{t}{imp} = 4.0\mu$s. The s-wave potential for the $\ket{\nu l} = \ket{133s}$ state is indicated without scale as gray lines for guidance.
Green vertical lines divide the radial coordinate into an inner region (I: $0\le r < R_{min}$) and an outer region (II: $r \ge R_{min}$). (b) The large density perturbations caused by the shape-resonance using the s+p-wave potential becomes more prominent in the outer region at times $\tau+\sub{t}{imag} = 15.0 \mu$s, after the ionization of the Rydberg impurity. Further snap shots of the density after free evolution are shown in (c) and (d) at $\tau+\sub{t}{imag} = 100 \mu$s and $\tau+\sub{t}{imag} = 231 \mu$s respectively. The inset of (c) shows the spatially averaged 3D density at $\tau+\sub{t}{imag} = 100 \mu$s.  The inset of (d) shows the variation of the averaged density over time at the Rydberg ion location ($x = y = z = 0$). For further details see the supplementary movie.
}
\end{figure}

To understand if and on what length-scales points (i) and (ii) make a difference, we compare the BEC dynamics for all three potentials in \fref{fig_single_comp}. For guidance, we divide the radial range into two regions, defining an inner (I: $0 \le r < \sub{R}{min}$), and an outer (II: $r \ge \sub{R}{min}$) region as indicated by the solid vertical green line as shown in \fref{fig_single_comp}. As evident from panel (a) the impact on the condensate is the same in the inner region for the s-wave and CASW potentials because these two potentials are near identical in this range, while in the outer region the s-wave response matches with that due to the s+p-wave interaction potential since here these other two potentials largely agree. In comparison to the s-wave potential and CASW, the s+p-wave potential causes a much stronger perturbation in the condensate density at the end of the imprint time, leading a relative increase of $\Delta \rho/\rho_0 \approx 10$ at the location of the ionic core. This is due the presence of the shape-resonance which provides a large central peak $V \approx -680$ MHz, see \fref{sketch}{c}, which was cut at $|\sub{V}{cut}| = 1$ MHz for the simulations in \fref{fig_single_comp}. Even when cutting off the s+p potential, it remains much stronger than that using only the s-wave approximation in the central region. 

As can be seen in \fref{fig_single_comp}{b}, for all three potentials the initial inward motion of the condensate is converted to an outward flow due to inertia at later times after the Rydberg impurity was removed, as discussed in \sref{single}. The wave created by the s+p-wave potential at $\tau + \sub{t}{imag} = 15.0\mu$s has reached the outer region and still has an about five times larger amplitude than that generated by the other two potentials. Therefore, a signature of the shape resonance should be accessible through very high resolution in-situ density measurements \cite{Wilson_insitu_vortex,electron_microscopy_BEC,Vestergaard_Hau_nearresimaging}. 

While the density waves created near the position where the Rydberg core has been removed depend on the level of detail used in the potential, we see that at much later times and larger radii, the waves created by all three potentials nearly agree. This shows that this part of the wave dynamics is not too sensitive to either fast oscillations or deviations between s-wave and s+p-wave potential in the inner region, see \fref{fig_single_comp}(c) and (d). We do, however, find that the smoothened relative density perturbations $\Delta(\delta\rho)$ caused by the s+p-wave potential remain approximately $1.5$ times ($2$ times) larger than those due to the s-wave potential (CASW potential), in the region $r< 1 $ $ \mu $m at $\tau + \sub{t}{imag} = 100\mu$s (see the inset of panel (c)).
Moreover, when directly comparing the average density due to the s-wave and CASW potential in the inset of the panel (c), one finds that the average maximum density ($\delta\rho$) with the s-wave potential is about $1.5$ times higher than from the CASW potential in the region $r<1$ $\mu$m at $\tau + \sub{t}{imag} = 100\mu$s. Thus, further out, the radial wells in the potentials play a slight quantitative but not a qualitative role in the Rydberg-BEC dynamics. In all cases atoms are first focussed inwards and only then diffuse outwards, as discussed in \sref{single}. This results in the non-monotonic behavior of the central density at the ionic core, visible in the inset of \fref{fig_single_comp}{d}.

An important conclusion that we can already draw from this present analysis is that the net condensate perturbation at large $R$ and late times $t$ is qualitatively insensitive to either of the details (i) and (ii) defined above. Both features are characterised by length scales much smaller than the healing length of $\xi=0.13$ $\mu$m, so that the condensate will only respond according to an averaged effect. This spatial averaging will make the impact of details even less prominent when the impurity is moving, as we have shown in \cite{tiwari:tracking}.


\subsection{Dependence on numerical potential cutoff $\sub{V}{cut}$}
\label{cutoff}

\begin{figure}
\includegraphics[width=0.8\columnwidth]{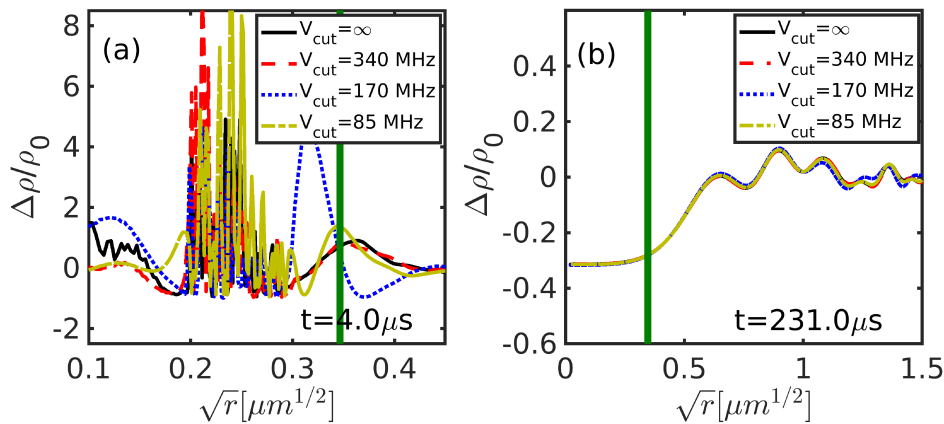}
\centering
\caption{\label{fig_cuts} Dependence of atom dynamics on potential cutoffs. (a) Relative condensate density variation   $\Delta \rho = \rho - \rho_0$ for $\rho_0 = 400 \mu$m$^{-3}$ in the inner region at the imprinting time $\sub{t}{exc} + \sub{t}{imp} = 4.0\mu$s. We use $\sub{V}{cut} = \infty$ (no cut-off, solid black lines), $\sub{V}{cut} = 340$ MHz (dotted red lines), $\sub{V}{cut} = 170$ MHz (dashed blue lines) and $\sub{V}{cut} = 85$ MHz (dot-dash yellow-sh lines). The resulting dynamics after removing the Rydberg atom, at later times and larger radii, is shown in (b), for $\tau + \sub{t}{imag} = 231.0\mu$s.  The solid green line at $r = \sub{R}{min}$ divides the potentials into two regions at the shape resonance.}
\end{figure}

While the s+p-wave potential agrees with the s-wave potential in the outer region, they significantly deviate close to the nucleus where the former reaches down to $V = -680$ MHz. This poses severe challenges for a wide range of schemes used to numerically propagate the 3D GPE, for which the potential drop enforces very small time steps. Here we thus separately analyze the importance of a complete inclusion of the potential step, using the radial GPE \bref{radialGPE}, cutting the central potential peak at $V = -\sub{V}{cut}$, as shown in \frefp{sketch}{c}, and then varying $\sub{V}{cut}$. 
Since the condensate response for the outer region at the imprint time $\sub{t}{exc} + \sub{t}{imp} = 4.0\mu$s will obviously still be the same for different cutoff values, we only show the zoom onto the inner region in \frefp{fig_cuts}{a}. 

As expected, the density perturbations of the condensate
in the inner region significantly depend on $\sub{V}{cut}$. However, counterintuitively, the amplitude of the perturbations does 
not follow a monotonic trend with $\sub{V}{cut}$. For the snapshot shown, the oscillations are largest for $\sub{V}{cut} = 85$ and $340$ MHz while they are significantly smaller for $160$ MHz and $\infty$ (no cutoff) in the region $0.2\mu$m$^{1/2}$ $< \sqrt{r} <$ $0.3 \mu$m$^{1/2}$. Relative amplitudes then vary in time. 
However, while the initial density fluctuations in the inner region of the potential depend strongly on the magnitude of $\sub{V}{cut}$, we find that the late time dynamics of the condensate becomes nearly cut-off independent, see \fref{fig_cuts}{b}. 

We can thus conclude that the huge central dip of the s+p-wave does not contribute significantly to the overall late time dynamics of the BEC. Therefore, we fix the cut-off at $|\sub{V}{cut}| = 2$ MHz for further investigation in the coming sections. 

\subsection{Repeated excitation}
\label{repeated}

In \sref{single} we found that a single Rydberg excitation only causes a minor relative change (a maximum of about $1\%$) of the atom density on percent level at $\tau + \sub{t}{imag} = 260\mu$s. As a consequence, experiments will likely require repeated excitations to measurably affect the bulk density. First steps in this direction have been recently taken in \rref{balewski:elecBEC}. Here, we model a sequence of $\sub{N}{exc}=10$ repeated excitations with a probabilistic approach. In contrast to the case of single excitation, the Rydberg atoms are no longer located at the origin, but have random positions with a Gaussian distribution using  
standard deviations $\sigma_x = 1.0 {\mu}$m, $\sigma_y = 2.0 {\mu}$m, and $\sigma_z = 1.0 {\mu}$m respectively. This assumes a relatively tight excitation laser focus as sketched in \fref{sketch}{a}. We have modelled a specific excitation scenario to obtain the spatial widths that we describe in 
 \aref{app_Rydexcitation}. The uncertainty in excitation location breaks spherical symmetry, so that we have now to turn the full solutions of the 3D GPE \bref{impurityGPE}.   
 For the temporal sequence we use an excitation duration of $\sub{t}{exc} = 0.5 {\mu}$s, imprint time $\sub{t}{imp} = 3.5 {\mu}$s and ionization time of $\sub{t}{ion} = 0.8 {\mu}$s. Thus $\sub{N}{exc}$ repeated excitations will require a total time $\sub{N}{exc}\times 4.8 \mu$s.
 
Since we are starting from an already quite dense condensate with $\rho_0 = 4\times 10^{14}$cm$^{-3}$, and find a large increase in condensate density around the core of the Rydberg atom during the time evolution see, e.g., \fref{fig_single_comp}, it is conceivable that three-body losses might become relevant. To investigate this, we have include the three-body loss term in the simulations for this section, and compare the total loss of atoms with and without Rydberg excitation.
While $500$ atoms are lost out of a total of $6.7\times 10^5$ atoms in our simulation volume, the difference between those two scenarios is negligible. Hence we conclude that Rydberg excitations do not significantly increase atom losse. This implies that the overdense region are too small to induce strongly enhanced three-body losses.

\begin{figure}
\includegraphics[width=0.8\columnwidth]{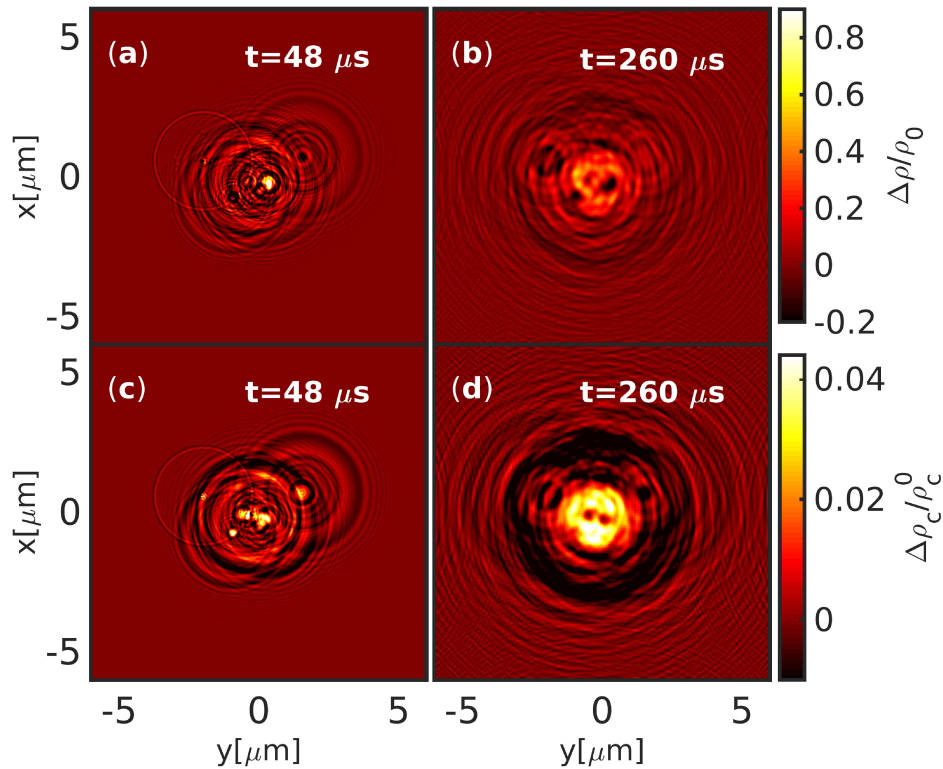}
\centering
\caption{\label{fig_repeat} Relative condensate density after repeated excitations of Rydberg impurities. (a) The 2D slice of the 3D density at $z = 0$ after $\sub{N}{exc} = 10$ repeated excitations at $t= \tau' = 10\times 4.8 = 48 {\mu}$s. The radius of circular features is that of the Rydberg orbit, which is about $1.8 \mu$m for $\ket{\nu l} = \ket{133S}$. The subsequent time evolution of the density is shown in (b) at $t=\tau' + \sub{t}{imag} = 260 {\mu}$s. The relative column density under the same conditions as in (a,b) is shown in (c,d). See also supplementary movie.
}
\end{figure}
We show the results after $\sub{N}{exc}=10$ excitations in \frefp{fig_repeat}{a}, using the s+p-wave interaction potential \bref{S_P_wave}. The figure shows 2D slices of the 3D density at $z = 0$, as well as column densities $\rho_c=\int dz\: \rho(\mathbf{R})$. Each circular feature is due to one Rydberg excitation, see supplementary movie. In contrast to the case of a single Rydberg atom in the BEC, right after the excitation a significant accumulated impact ($\Delta(\delta\rho)/\rho_0 \approx 1 \%$) can be seen in the smoothened condensate column density after $10$ repeated excitations are completed. 
The impact becomes even more evident after an additional evolution time of $\tau' + \sub{t}{imag} = 260 {\mu}s$ as shown in \fref{fig_repeat}(b), where $\tau' = 10\times (\sub{t}{exc} + \sub{t}{imp} + \sub{t}{ion}) = 48 {\mu}s$ is the time taken to complete 10 repeated excitations with the subsequent removal of the Rydberg exitation.

\begin{figure}
\includegraphics[width=\columnwidth]{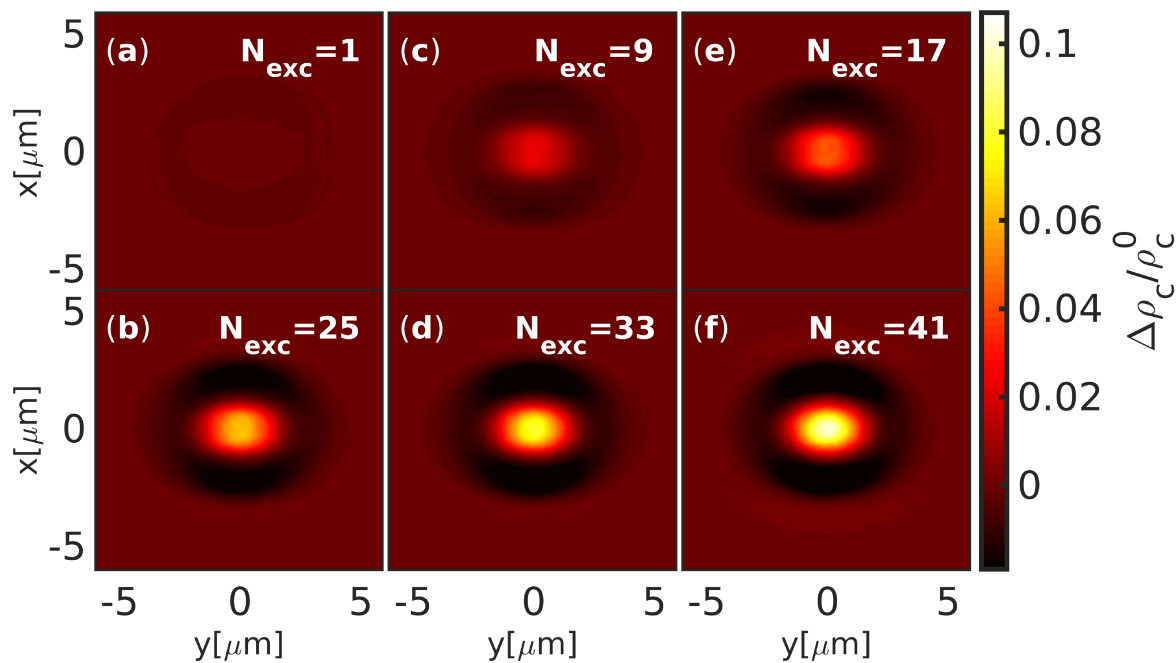}
\centering
\caption{\label{fig_pulses} Signals using $\sub{N}{exc}$ repeated Rydberg excitations. The column density relative to the background column density $\rho^0_{c} = 4.76 \times 10^{3} {{\mu}m}^{-2}$, i.e. $\Delta\rho_c = \rho_c - \rho^0_c $ is shown here. (a) Column density for a single Rydberg excitation ($N_{exc}$ = 1) at $\tau + \sub{t}{imag} = 260 \mu$s, averaged over $400$ samples of Rydberg positions. As the number of repeated excitations is increased from $\sub{N}{exc} = 1$ to $\sub{N}{exc}=41$, the features in the column density become more pronounced, from (a) to (f).
}
\end{figure}

While Rydberg excitations leave a visible mark in both densities
a detailed inspection of the contrast between maximum and minimum, reveals a higher maximal contrast in the 3D density ($8\%$ of the mean) than the column density ($1\%$ of the mean), even though the simulation box is only extended over $L_z=6$ $\mu$m in the $z$-direction.
As part of a larger cloud, the column density signal would be even weaker.
Nonetheless, for the case here, the signal appears prominent over a larger spatial region in the column density, leaving a clear density depression of $1.2\%$ in panel (d) around the central enhancement of the density. The latter is due to the constructive interference of waves directed towards the centre by each of the excitations, as discussed in \sref{single}.

An experimental density measurement might additionally involve an averaging over $\sub{N}{samp}$ repeated measurement. We explore how the expected density profile depends on the number $\sub{N}{exc}$ of repeated excitations in \fref{fig_pulses}. The figure shows the averaged column density at $t=4.8\times\sub{N}{exc} + \sub{t}{imag} = 260 {\mu}s$, subtracted from the background column density at $t = 0$, and averaged over $\sub{N}{samp}=400$ samples, each with a different random realisation of the Rydberg atom positions. We show each 
snapshot at the same time $t$, hence the delay $\sub{t}{imag}$ between the last excitation and the snapshot shown 
is varying as $\sub{t}{imag} = (260 -4.8\times\sub{N}{exc})\mu$s.  
We clearly see that by increasing the number of excitations from $\sub{N}{exc}=1$ to $\sub{N}{exc}=40$, the relative density contrast of the resultant feature improves from about $0.1\%$ to  near $11\%$. As before, we see that the net effect is a central density increase, surrounded by a density depression. The feature should be visible through in-situ density measurements \cite{Wilson_insitu_vortex,electron_microscopy_BEC,Vestergaard_Hau_nearresimaging}.  
\begin{figure}
\includegraphics[width=0.8\columnwidth]{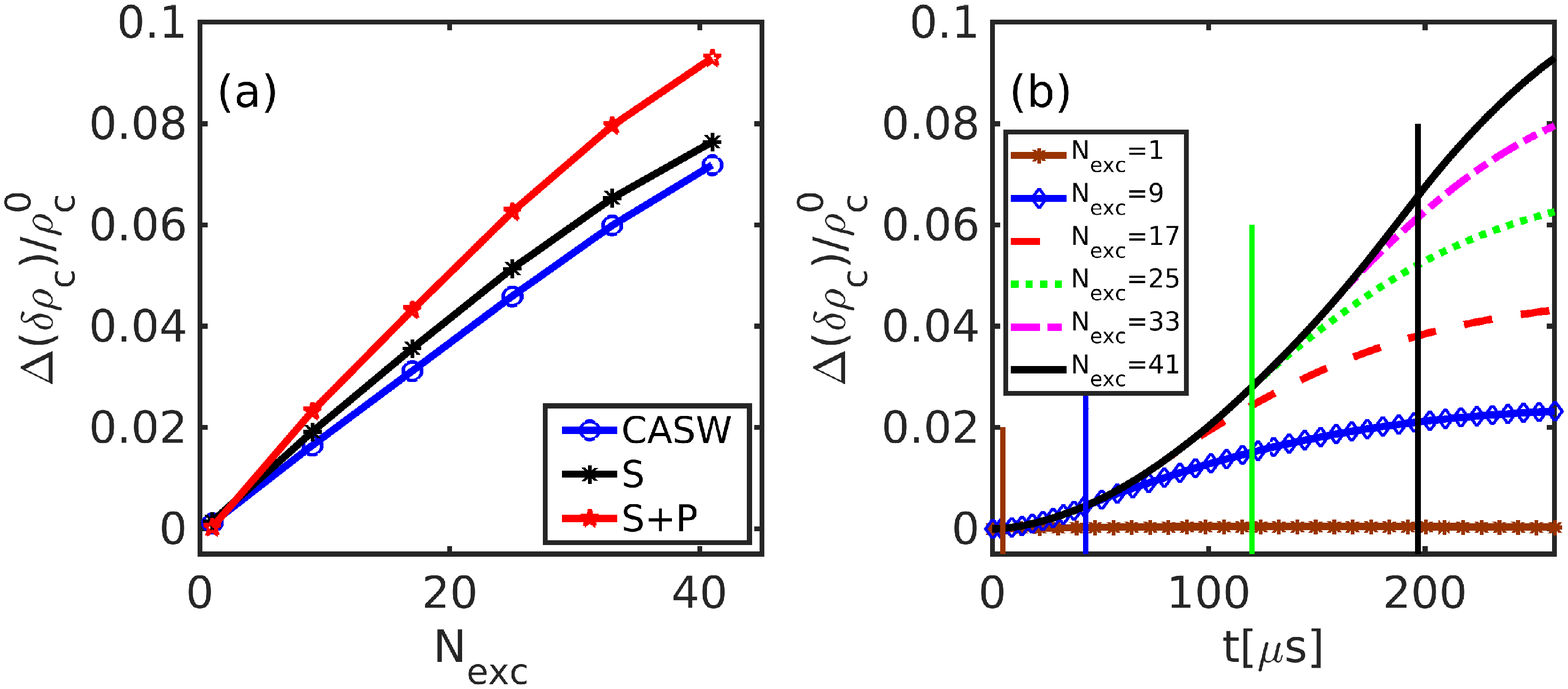}
\centering
\caption{\label{Diff_Pulses} Increase in the smoothened relative column density at the ion location ($x = y = 0$) for different repeated excitations comparing in (a) the s-wave potential \eref{Swave} (black solid and $*$), CASW potential \eref{Classpotfull} (blue solid and $\circ$) and the s+p-wave potential \eref{S_P_wave} (red solid and $\star$), at a fixed time $t=\sub{N}{exc}\times 4.8\mu$s + $\sub{t}{imag} = 260\mu$s. (b) Evolution of the smoothened column density for different numbers of repeated excitations at $x = y = 0$, considering the s+p-wave interaction potential. Smoothing uses a  Gaussian kernel with a standard deviation of $\sigma = 0.5 \mu$m in both panels. Vertical colored lines indicate at what time a given number of excitations $\sub{N}{exc}$ is complete.
}    
\end{figure} 
Next, we compare the increase of the excess in the average column density at the centre at $x = y = 0$ through the repeated excitations, using the three different potentials. The results are shown in \fref{Diff_Pulses}(a), averaged over $\sub{N}{samp} = 400$ samples. The data clearly follows the same pattern for all three potentials, but with slightly different slopes. While the full s+p-wave potential creates the largest density increase, a difference is still visible between the signature generated by the s-wave and CASW potentials, which again suggests that the radial wells do not qualitatively change the picture but have a discernible quantitative effect.

Note, that we have restricted our simulation to $128\times128\times128$ grid points which allowed us to afford the average over $\sub{N}{samp} = 400$ position realisation. Based on our findings in \sref{single}, we expect this to undersample the interaction potential, but not lead to a 
 qualitatively modified condensate dynamics. Nonetheless, the net density contrast for the s-wave and s+p-wave potentials will increase by up to a factor of two for more grid points $512\times512\times512$. Therefore, due to computational constraints, we report a lower bound on the density perturbation caused by repeated excitations.
 
Finally, in \frefp{Diff_Pulses}{b} we show the increase in time of the signal contrast, for different numbers of excitation $\sub{N}{exc}$.
 We can see that the largest contrast that is reached at late times scales roughly linear with $\sub{N}{exc}$, while the earliest time for which a given fixed contrast is reached no longer reduces with $\sub{N}{exc}$ after a certain number of excitations. For example, a contrast of $2\%$ is reached at $t\approx 100\mu$s in \fref{Diff_Pulses}, for all $\sub{N}{exc}\geq17$ regardless of $\sub{N}{exc}$. We see that for $\sub{N}{exc}>1$, an additional time of Rydberg-free evolution significantly enhances the signal, and thus might be an important experimental tool for a more visible effect. 

\section{Local heating of the condensate}
\label{TWA} 
%
It has been found previously that a Rydberg impurity excited in the BEC causes atom-loss and heating, increasing with the number of repeated excitations of a Rydberg impurity \cite{balewski:elecBEC,Amartin}. In two-dimensions, we have shown in \rref{tiwari:tracking} that heating is limited when a small number of impurities is excited in a low density BEC background within the time span of a few microseconds. Specifically, the number of atoms entering the uncondensed fraction is not enough to invalidate the GPE.
In this section we show that these conclusions remain unchanged for the scenarios considered in the work, i.e. in 3D and at higher densities. Importantly, we find that beyond mean field physics also do not significantly affect any of the earlier results discussed in this article. As we found that the three-body loss is negligible we do not consider its contribution in this section. 

To study heating in the present context, we employ the truncated Wigner approximation (TWA) \cite{steel:wigner,Sinatra2001,castin:validity,wuester:nova2,wuester:kerr,wuester:collsoll,norrie_wignerK3} which extends 
 \erefs{impurityGPE}{GPE} beyond mean-field theory. This is done by adding quantum and thermal fluctuations to the initial state through a specific recipe by the inclusion of random noise. The 
 total atomic density $\rho_t$ can then be split into the condensed density $\rho_c$ and the uncondensed one $\rho_u$, after averaging over an ensemble of $\sub{N}{traj}$ realisations of the noisy simulations as discussed in \aref{app_TWA}. 
For a single excitation with s+p-wave interaction potential \bref{S_P_wave} and averaging over random impurity positions as in \sref{repeated}, we find that until $\sub{t}{exc} + \sub{t}{imp} = 4.0$ $\mu$s, while the Rydberg impurity is present, it causes only an additional $\approx 65$ atoms to become uncondensed, compared to their initial number, out of a total of $6.7\times 10^{5}$ atoms in our simulation box.
Focussing exclusively on the Rydberg excitation volume, a sphere with radius $2\mu$m, the uncondensed number thus increases by about $50$ out of $13400$ atoms, corresponding to $0.4$\%. When adopting the repeated excitation scheme described in \sref{sequence} under the conditions of \fref{fig_repeat} ($\sub{N}{exc} = 10$), we find that about $1300$ atoms in the box, and roughly $1200$ ($9$\%) of the atoms in the Rydberg excitation volume are depleted from the condensate, compared to the initial state, up to time $t=10\times(\sub{t}{exc} +\sub{t}{imp}) +9\times \sub{t}{ion} = 47.2$ $\mu$s. These results suggest that heating is not strong enough to significantly alter the results of our earlier mean-field simulations, as we indeed shall see shortly.

\begin{figure}
\includegraphics[width=0.9\columnwidth]{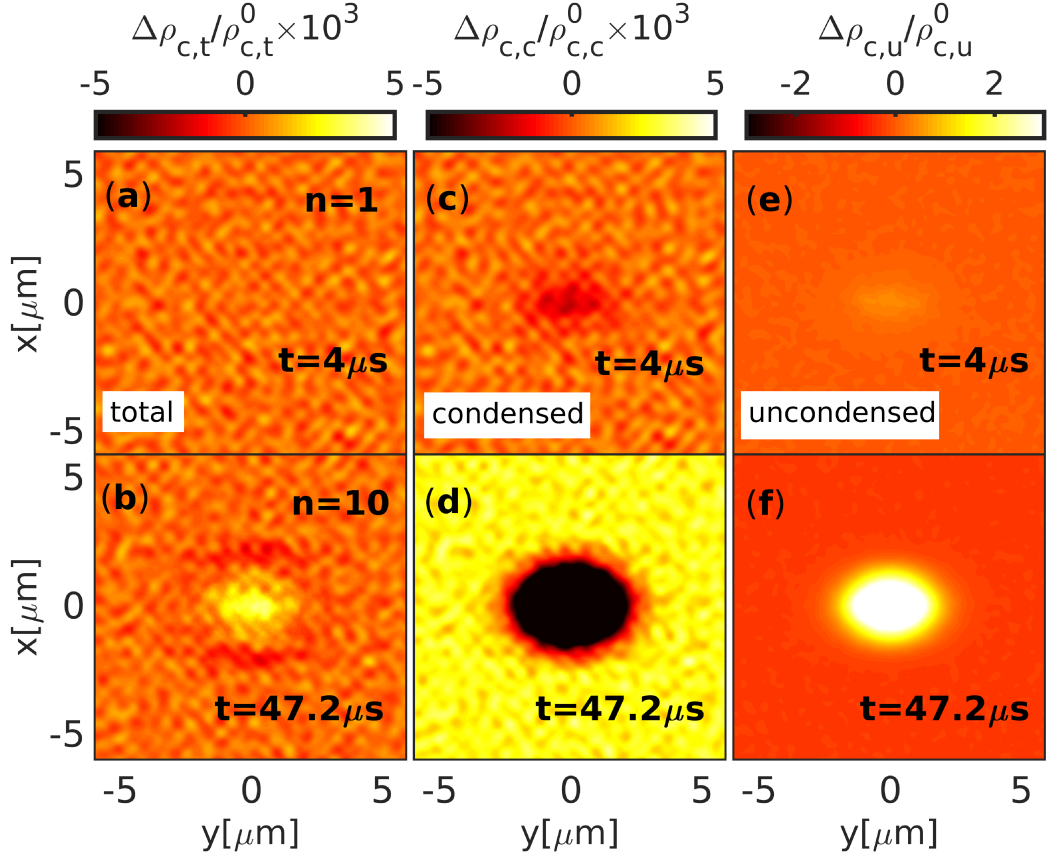}
\centering
\caption{\label{fig_singleTWA} Beyond mean-field response of the Bose gas, using the s+p-wave potential \bref{S_P_wave}, and averaging over $\sub{N}{traj}=5120$ trajectories for a single (top row) and ten (bottom row) excitations. The total relative column density, $\sub{\rho}{c,t}$, for a single (a) and ten repeated excitations (b) at $\sub{t}{exc} + \sub{t}{imp} = 4.0 {\mu}$s and $10\times(\sub{t}{exc} +\sub{t}{imp}) +9\times \sub{t}{ion} = 47.2 {\mu}$s shows a weak impact of the Rydberg impurities within their respective imprint time, where $\rho^0_{c,t} = 4.7\times 10^{3} {\mu m}^{-3}$ is the initial total column density. A clearer local signal can be seen in the condensed (d) and uncondensed components (f) of the BEC for ten repeated excitations, and a faint signal for the single excitation in the condensed (c) and uncondensed component (e),
 where $\rho^0_{c,c} = 4.72\times10^{3}{\mu m}^{-3}$ and $\rho^0_{c,u} = 2.61\times10^{1}{\mu m}^{-3}$ are the initial condensed and uncondensed column densities.
We use a homogeneous mean density of $400.0 {\mu m}^{-3}$.
}
\end{figure}

When moving from bulk-data such as uncondensed atom numbers to local data such as the uncondensed density $\sub{\rho}{u}(\bv{x})$, we see that the condensate heating is restricted to within the excitation volume. This is demonstrated in \fref{fig_singleTWA}, for the conditions of the single and ten repeated excitations.

The first row contains the relative change of the total column density, $\rho_{c,t}$, condensed column density, $\rho_{c,c}$ and uncondensed column density $\rho_{c,u}$ after a completed single excitation, at $ \sub{t}{exc} + \sub{t}{imp} = 4.0 {\mu}$s. The components are extracted from the stochastic field as described in \aref{app_TWA}. The second row shows snapshots after $\sub{N}{exc} = 10$ repeated excitations at $t=10\times(\sub{t}{exc} +\sub{t}{imp}) +9\times \sub{t}{ion} = 47.2 {\mu}$s. It is evident from the uncondensed density in panel (e) that heating remains confined to within sphere with radius of $2 \mu$m around $x = y = 0$ for a single excitation. This in turn leaves a hole of relative depth of about $0.2$ \% in the same region of the condensed column density, see \fref{fig_singleTWA}{c}. However, since overall no atoms are lost, no features are caused at this time in the total density shown in \fref{fig_singleTWA}{a}.  The depth of the dip in the condensed density increases by about $ 6.0 \%$ as one moves from a single to ten repeated excitations, as shown in \fref{fig_singleTWA} (d). This happens since more atoms transfer to the uncondensed component for ten repeated excitations in comparison to a single excitation. Comparing the condensate density from TWA as shown in \frefp{fig_singleTWA}{d}, see also \aref{app_TWA}, with the condensate density using the GPE (not shown) we find a qualitatively different behaviour as expected, since the GPE cannot describe local heating. Note, however, that experiments would only measure the total density as shown in \frefp{fig_singleTWA}{b} that appears similar to the mean field result.

\begin{figure}
\includegraphics[width=0.9\columnwidth]{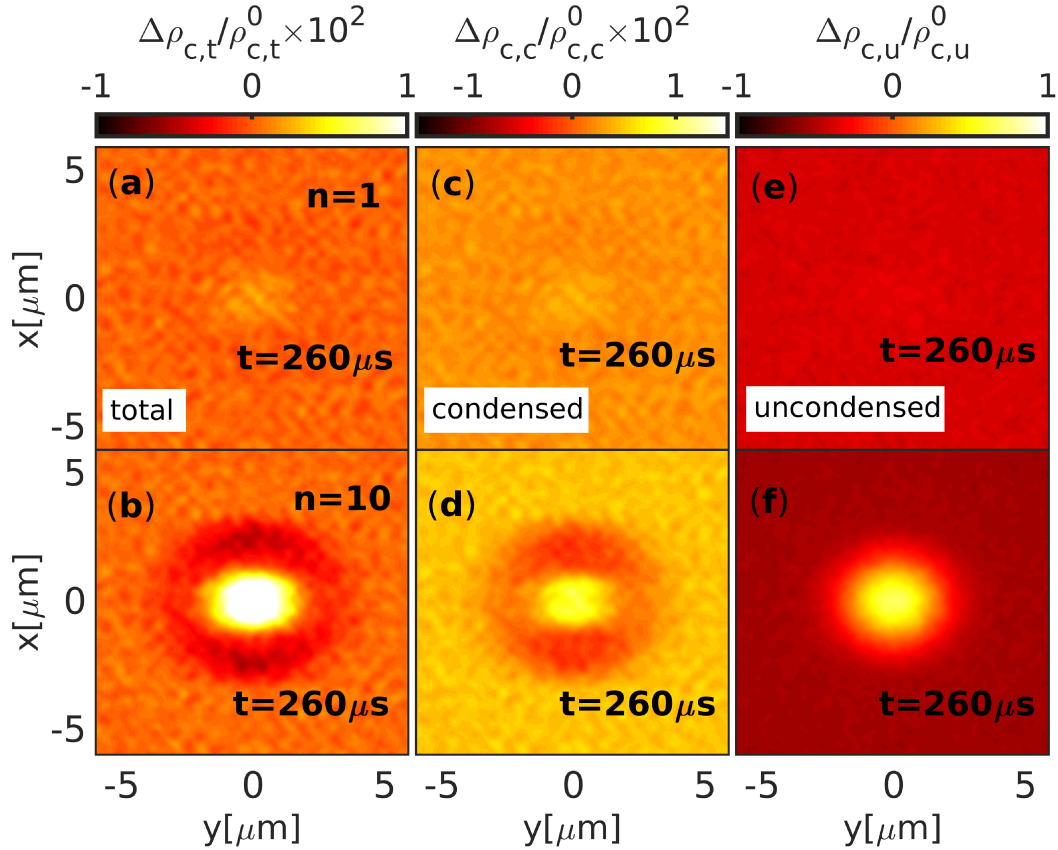}
\centering
\caption{\label{singleTWA_sup} Evolution of the Bose gas following the Rydberg excitation beyond mean-field theory, for the same parameters as in \fref{fig_singleTWA}. (a,b) The total column density for a single and ten repeated excitations at $t=\tau + \sub{t}{imag} = 260.0$ $\mu$s and $t=\tau' + \sub{t}{imag} = 260$ $\mu$s respectively shows a significant impact of the Rydberg impurities. We also show the condensed column density
after a single excitation (c) and after ten repeated excitations (d), as well as the uncondensed  column density after a single excitation (e) and after ten repeated excitations (f). Note the different colorscale compared to \fref{fig_singleTWA}.
}
\end{figure}
We can see in \frefp{fig_singleTWA}{b} that a faint signature is visible already after $\sub{N}{exc} = 10$ repeated excitations in the total density directly after the excitation period, despite the added noise due to residual fluctuations in the mean. The contrast is, however, comparable to the one in the mean-field scenario shown in \frefp{fig_repeat}{c}, with the difference originating in the different number of realisations in the simulation.
By waiting an additional time the signal to noise ratio can be further enhanced, as suggested by \frefp{Diff_Pulses}{b}. We show this explicitly in \frefp{singleTWA_sup}{b} after a free time evolution of $t=\tau' + \sub{t}{imag} = 260\mu$s. The detailed time evolution of the local heating dynamics for single and ten repeated excitations is shown in the last two column of \fref{singleTWA_sup}, indicating significant motional dynamics also in the uncondensed component at later times. If we now compare the late-time condensate density in TWA, \frefp{singleTWA_sup}{d}, with the corresponding one using the GPE, \frefp{fig_pulses}{c} shows that these largely agree, 
as expected based on our earlier observation that only a small fraction of atoms becomes uncondensed and
hence BEC bulk dynamics should be well described by the GPE simulations.

\begin{figure}
\includegraphics[width=0.8\columnwidth]{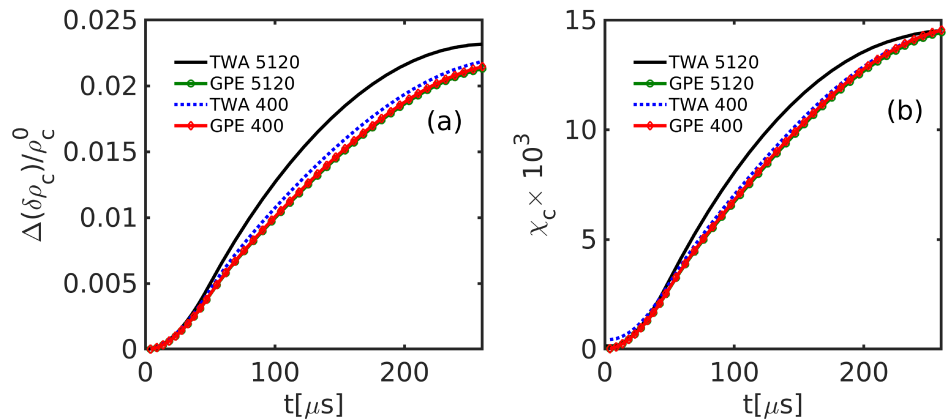}
\centering
\caption{\label{TWA_GPE} The column density averaged over a Gaussian kernel with width $\sigma = 1 \mu$m
is compared in (a) at the Rydberg core between mean-field simulations for $\sub{N}{traj} = 5120$ (solid green and o) and $\sub{N}{traj} = 400$ (solid  red and $\Diamond$) and TWA simulations for $\sub{N}{traj} = 5120$ (solid black lines) and $\sub{N}{traj} = 400$ (dotted blue lines). The contrast \bref{contrast} of the same quantities is shown in (b).
We use ten repeated excitations, as in \fref{fig_singleTWA}. The initial background column density for both simulations is $\rho^0_c = 4.76 \times 10^3$ $\mu$m$^{-2}$. }    
\end{figure}

For a more quantitative comparison between TWA and GPE, we define the image contrast $\chi_c$ as the difference between the maximum and minimum in the smoothened relative column density according to
\begin{eqnarray}
\chi_c &= \frac{1}{2}\left[\bigg(\frac{\Delta(\delta\rho_c)}{\rho_c^0}\bigg)_{\text{max}} - \bigg(\frac{\Delta(\delta\rho_c)}{\rho_c^0}\bigg)_{\text{min}}\right],
\label{contrast}
\end{eqnarray}
where $\delta\rho_c$ is the smoothened column density, and $\Delta\delta\rho_c$ thus $\delta\rho_c - \rho_c^0$ .
In \fref{TWA_GPE} we compare the time evolution of $\chi_c$ obtained from GPE simulations averaged over $\sub{N}{traj} $ different realisations of atomic positions, with TWA simulations of $\sub{N}{traj}$ trajectories. The latter combines two different averages: in each trajectory the atomic positions as well as the quantum noise realisation are different. The figure then compares $\sub{N}{traj} =400$ and $5120$ trajectories. For better comparability, all column densities are first smoothened with a Gaussian kernel of resolution $1.0$ $\mu$m.

We see that the difference between TWA and GPE is very small when averaging over the same number of impurity positions, as expected from our earlier observation of only a minor impact of condensate heating. Increasing the sample size maintains the maximum contrast but results in a smoother image. Since this applies for the GPE simulations as well, we conclude that the average over all spatial realisations is converged at the chosen number of trajectories $\sub{N}{traj}$. For our 3D TWA calculations of a single and ten repeated excitations, we employed $128 \times 128 \times 128$ spatial grid-points and averaged over $\sub{N}{traj} = 5120$ trajectories, which results in about $0.5\%$ standard error on mean 3D densities.
%
%
\section{Conclusions}
\label{conclusion}
We presented an extensive campaign of simulations of Bose-Einstein condensate dynamics in response to the excitation of multiple or single Rydberg atoms. We have described  the quantum dynamics across all relevant length scales, ranging from microscopic scales within individual nanometer-sized radial wells of the Rydberg-BEC interaction potential, out to mesoscopic distances of twice the Rydberg orbital radius. We find that, while the dynamics within the Rydberg orbit is sensitive to the level of detail used to describe that interaction potential, density perturbations travelling outwards at later times are well reproduced also when using approximate interaction potentials, dramatically simplifying simulations. 

We explored the shape and contrast of density perturbations as the number of Rydberg insertions is increased, and found that the typical response is the transient development of a density depression surrounding a density increase at the centre of the Rydberg excitation volume. The contrast of this feature grows linearly with the number of Rydberg insertions. While signal in condensate density created by a single Rydberg atom remains very weak, it steadily increases if atoms are excited repeatedly. Importantly, we find that even after cessation of the excitation sequence, the signal increases further for a while as outgoing density waves combine. This suggests an additional unperturbed evolution interval as a helpful experimental tool to allow the Rydberg excitations to strongly affect the BEC.

Finally we went beyond mean-field theory, adding quantum fluctuations and the possibility of atoms being ejected from the condensate, using the truncated Wigner approximation \cite{steel:wigner,Sinatra2001,castin:validity,wuester:nova2,wuester:kerr,wuester:collsoll,norrie_wignerK3}. We find that a Rydberg impurity excited in the BEC causes additional uncondensed atoms, largely concentrated in the excitation region.
The number of uncondensed atoms increases during a sequence of repeated excitations of a Rydberg impurity, but does not become large enough to invalidate the use of mean-field simulations and hence any of the conclusions listed above.

While we find that small scale details of the Rydberg-BEC potential, such as the p-wave shape resonance and radial oscillations, do not qualitatively affect the BEC response, larger features such as the anisotropy of $p$ or $d$ states \cite{Karpiuk_imaging_NJP} should leave a discernible signature after repeated excitations. 
Also different condensate atom species might give valuable information, such as $Sr$ for which the shape resonance is absent, and $Cs$, the different mass of which
will alter kinematics. All these should be subject of further investigations.

\ack
S.T.~and S.W.~thank the Max-Planck society for financial support through the MPG-IISER partner group program. F.~M.~acknowledges funding from Deutsche Forschungsgemeinschaft [Project No. PF 381/17-2, part of the SPP 1929 (GiRyd)], the Carl Zeiss Foundation via IQST, and is indebted to the Baden-W{\"u}rttemberg-Stiftung for the financial support by the Eliteprogramm for Postdocs. R.~S.~and M.~W.~acknowledge support by the Max Planck Society, the Deutsche Forschungsgemeinschaft (DFG, German Research Foundation) under Germany€'s Excellence Strategy €"EXC-2111-390814868 and within the priority program ``€œGiant interactions in Rydberg systems'', DFG SPP 1929 GiRyd, Grant No.~428462134. 

%
\appendix

\section{Rydberg excitation locations}
\label{app_Rydexcitation}
%
The location of a Rydberg excitation in a dense BEC can be controlled by a combination of a tightly focussed excitation beam
and the Rydberg state energy dependence on background density \cite{balewski:elecBEC}.
To obtain an experimentally relevant spatial distribution of Rydberg excitations within a harmonically confined BEC, we model the excitation process taking these two features into account.

We start by generating clouds of point-like particles with a distribution function matching the Thomas-Fermi profile of
of the BEC density $\rho(\mathbf{R})$ in a cigar shaped harmonic trap, with long axis along $y$. In cylindrical coordinates
\begin{eqnarray}
\rho(r,y,\phi) =  \frac{1}{U_0} \left( \mu - \frac{m}{2} (\omega_r^2 r^2 + \omega_y^2 y^2) \right)
\times \Theta \left(  \mu - \frac{m}{2} (\omega_r^2 r^2 + \omega_y^2 y^2) \right),
\label{TFprofile}
\end{eqnarray}
where $\Theta$ is a Heaviside step function, $r = \sqrt{x^2+z^2}$ (in this appendix only), $\mu$ the chemical potential and we assume trapping frequencies $\omega_r$ in the radial direction and $\omega_y$ along the long axis of the BEC. Note, that the Thomas Fermi profile is used in this appendix only, while the main article assumes a homogenous background BEC. 

From the density distribution we can derive the corresponding cumulative distribution functions (CDFs) for each coordinate that allow us to randomly draw particle positions matching the probability to find atoms at a certain position inside the BEC. The extension of the BEC in the long direction is between $y_{\text{max}}=-y_{\text{min}}= \sqrt{2\mu/(m \omega_y^2)}$, and then for
each $y$ an upper bound for the radial coordinate is given by $r_{\text{max}}(y)=1/\omega_r \sqrt{2\mu/m- \omega_y^2 y^2}$. Consequently, 
the total particle number of the BEC is given by
\begin{eqnarray}
N = &  \int\limits_{y_{\text{min}}}^{y_{\text{max}}} \text{d} y \underbrace{\int\limits_{0}^{r_{\text{max}} (y)} \text{d} r \underbrace{\int\limits_{0}^{2 \pi} \text{d}\phi \ r\, n(r,y,\phi)}_{=\bar{n}_\phi(r,y)}}_{=\bar{n}_{r,\phi}(y)},
\end{eqnarray}
where $\bar{n}_{r,\phi}(y)$\ d$y$ gives the weight of a disc with thickness d$y$ (i.e.\ a cross-section of the BEC) and $\bar{n}_\phi(r,y)$\ d$y$ d$r$ characterizes the weight of a infinitesimal cylinder of radius $r$ for a given value of $y$.
Based on the latter, the CDF of the $y$-coordinate is given by 
\begin{eqnarray}
\text{CDF}_{{n}_{r,\phi}}(y) = \frac{1}{N}\,\int\limits_{y_{\text{min}}}^{y} \text{d} y^{\prime} \, \bar{n}_{r,\phi}(y^{\prime}).
\end{eqnarray}
It maps the allowed range of $y$ onto the interval $[0,1]$. Hence we can use the inverse CDF to draw random numbers $\xi^\ast_y \in [0,1]$ and assign them to $y$-coordinates $y^{\ast}$, which are then correctly distributed according to the Thomas-Fermi profile of the BEC.
Analogously, we obtain the CDF of the radial coordinate
\begin{eqnarray}
\text{CDF}_{\bar{n}_\phi}(r,y)\big|_{y=y^{\ast}} = \frac{1}{N} \, \int\limits_{0}^{r} \text{d} r^{\prime} \, \bar{n}_\phi(r^{\prime},y)\big|_{y=y^{\ast}}.
\end{eqnarray}
The inverse of $\text{CDF}_{\bar{n}_\phi}(r,y^{\ast})$ assigns a random variable $\xi^{\ast}_r \in [0,1]$ to an $r$-coordinate $r^{\ast}$ under the condition that the atom is found at the $y$-coordinate $y^{\ast}$, which has been obtained in the previous step. Finally the $\phi$-coordinate is drawn uniformly from the interval $[0,2\pi)$, as the density profile \bref{TFprofile} of the BEC is invariant under rotation around the $y$-axis. By successively drawing random coordinates as described above, we obtain 3D atom positions forming a cloud that matches the density distribution of a BEC confined in a harmonic trap. In the following we describe how to select atoms from this cloud to be excited into a Rydberg state by a Gaussian laser beam with a Gaussian profile and exploiting the density detuning of the energy of the Rydberg state.

The overall excitation probability is proportional to the intensity profile of the Gaussian beam which propagates anti-parallel to the $x$-axis of our coordinate system, see \frefp{sketch}{a}. To implement this, we evaluate the laser intensity profile
\begin{equation}\label{eq:App.A:intensity}
I(r_\perp, x) = I_0 \frac{\omega_0^2}{\omega(x)^2} \,e^{-\frac{2r_\perp^2}{\omega(x)^2}}
\end{equation}
at each atom within the cloud, where $r_\perp = \sqrt{y^2+z^2}$, $\omega(x)=\omega_0 \, \sqrt{1+x/x_R}$ and $x_R=\pi/\lambda \, \omega_0^2$ \cite{laserbook}. 
In accordance with a typical experimental setup, we choose $\omega_0 = 1.8 \, \mu m$ and $\lambda = 1.011 \, \mu m$. We keep $I_0$ dimensionless and determine it such that the sum of the $I=(r_\perp, x)$ at all atom positions is normalized to one. Assuming the excitation of exactly one Rydberg atom, the function $I$ can then directly be taken as the excitation probability of an atom at location $(r_\perp, x)$. This allows to select atoms to be excited into a Rydberg state according to the Gaussian intensity profile of the excitation laser.

In addition the dependence on light intensity, the excitation probability of an atom into a Rydberg state is density selective. This is because the spectral width of the laser only allows for excitations within a certain energy range $\Delta E$, and interactions between the Rydberg atom and the BEC, as in \sref{electron_atom}, cause an energy shift of the Rydberg state. We assume a Gaussian line shape of the excitation laser and hence a dependence of the excitation probability on energy shift $E$ as
\begin{equation} \label{eq:App.A:detuning}
p(E) = {\cal N}\, e^{-\frac{(E-\bar{E})^2}{\Delta E^2}},
\end{equation}
with $\Delta E=1$ MHz centered around a detuning $\bar{E}=-55$ MHz. 
If the normalisation factor ${\cal N}$ is chosen such that $p(-55\, \text{MHz}) =1$, this assigns a probability to each excitation with energy detuning $E$. For each atom, $E$ is found by summing the potential energy shift due to the potential \bref{S_P_wave} for all other atoms in the cloud:
\begin{equation}
p(E) = \sum_k \sub{V}{ryd,S+P}(\bv{R}_k).
\end{equation}
Here $\bv{R}_k$ is the location of the $k$th cloud atom and we set $\eta(t)=1$ and $\sub{\bv{x}}{n(t)}$ to the location of the atom for which we wish to evaluate the Rydberg excitation probability.

Finally the total Rydberg excitation probability for each atom in the cloud is given by $I(r_\perp, x)\times p(E)$. The histogram of excitation positions is well fitted by a three-dimensional Gaussian distribution, with the widths $\sigma_{x,y,z}$ given in the main text.
%
\section{Rydberg molecular potentials}
\label{app_Rydmol}
%
Since the interaction potential between the condensate and Rydberg impurity is solely governed by the electron-atom collision energy, the Fermi pseudo-potential \bref{Swave} is a valid approximation for the low energy Rydberg electron. As the electron gets closer to the ionic core, it gains more kinetic energy from the Coulomb potential, that eventually matches with energy of a quasi-bound Rb$^-$ state behind the p-wave centrifugal barrier \cite{Fabrikant_1986}. This causes a shape resonance in the scattering cross-section between electrons and ${}^{87}$Rb atoms in the ${}^{3}P^{0}$ scattering channel at $0.02 $eV \cite{Fabrikant_1986}. As a result, the Fermi pseudo-potential needs to be extended to include p-wave scattering terms \cite{Omont:Pwave} as in
\begin{eqnarray}
&\sub{W}{ryd,s+p,n}(\bv{R},\bv{r},t) = \eta(t)\bigg[V_0(\bv{R}) \delta^{(3)}(\bv{R}-\bv{x}_n(t) -\bv{r}) 
\CR
&+ \frac{6 \pi \hbar^2 a_p[k(\bv{R})]}{m_e} \delta^{(3)}(\bv{R}-\bv{x}_n(t) -\bv{r}) \overleftarrow{\boldsymbol{\nabla}}_{\bv{r}} {\cdot} \overrightarrow{\boldsymbol{\nabla}}_{\bv{r}}  \bigg],
\label{SPWave}
\end{eqnarray}
where $\bv{R}$ is the position of the ground state atom, $\bv{r}$ is the position of the Rydberg electron relative to the Rydberg core, and the latter is located at $\bv{x}_n(t)$. $\eta(t) = 1$ encapsulates the presence or absence of a
 Rydberg impurity in the BEC, and the first term in the square bracket is the usual s-wave pseudo-potential \cite{Fermi:Pseudo}, which upon taking the expectation value in the Rydberg state in the absence of the p-wave scattering term, results in the effective mean-pseudo-potential \bref{Swave} defined in section \ref{electron_atom}. The last line of \bref{SPWave} is the p-wave scattering term, with p-wave scattering volume $a_p[k(\bv{R})] = -\tan(\delta^p[k(\bv{R})])/k(\bv{R})^3$, where $\delta^p[k(\bv{R})]$ denotes the triplet p-wave scattering phase shift of $e^-$-Rb$^{87}(5S)$ \cite{Matt:Phase}.
In order to calculate the full Born-Oppenheimer potential energy surfaces of Rydberg electron-atom interaction from \bref{SPWave}, we apply degenerate perturbation theory, diagonalising the Rydberg atom Hamiltonian as in \bref{S_P_wave} including the interaction \bref{SPWave} as a function of distance $r=|\mathbf{R}|$ between perturber atom and Rydberg ion. For each diagonalization, a total of six different $\nu$ manifolds and their respective angular momentum states have been taken into account, where one of them is above the target state and five are below. The angular momentum states for $l\le2$ are calculated using Numerov's algorithm to account for the quantum defect \cite{Gallagher:Qdefect1,Gallagher:Qdefect2}, whereas hydrogenic basis states are utilized for higher angular momentum states. The extracted potential energy for $\ket{\nu l} = \ket{133S}$ is compared in \frefp{sketch}{c} with the more basic s-wave scattering potential \bref{Swave}.

As discussed in \eref{Classpotfull}, for a simpler Rydberg-ground-state potential, we also make use of the classical electron probability distribution \cite{Amartin} given by
\begin{eqnarray}
\rho^{\mbox{cl}}(\bv{R})=\frac{1}{8\pi^2 r}\frac{1}{\sqrt{\epsilon^2b^2-(r-b)^2}},
\label{Classdens}
\end{eqnarray}
where $r=|\mathbf{R}|$, $b=-k/2E$ is the semi-major axis for the elliptical electron orbit in a  Coulomb field $U(\bv{R})=-k/{r}$ with $E$ the energy of the $\nu^{th}$ level and $\epsilon=\sqrt{1+2EL^2/m_e k^2}$ the eccentricity. Here $L$ is the angular momentum of the Rydberg state and $m_e$ is the mass of the electron.  
%
\section{Radial Gross-Pitaevskii equation in a homogeneous system}
\label{app_radial}
%
In \sref{single} we need to solve the radial GPE \bref{radialGPE} to fully resolve the many small scale oscillations of the Rydberg-molecular
potential. We wish to evaluate the derivatives in the GPE using the Fast-Fourier-Transform algorithm (FFT), which implicitly enforces spatial periodicity. This is not straightforwardly possibly for a radial coordinate and a homogenous BEC background, but requires the tools discussed in this apperndix.
 
We firstly work with a shifted radial wavefunction
\begin{eqnarray}
\tilde{u}(r,t) = u(r,t) - r\sqrt{\rho}e^{-i\mu t/\hbar}.
\end{eqnarray}
that is designed such that $\tilde{u}(r,t)=0$ at large $r$, where the background is unperturbed.  In the above, $\mu=U_0 \rho_0$ is the chemical potential.

The new radial GPE for the variable $\tilde{u}(r,t)$ is then:
\begin{eqnarray}
\label{OffsetGPE}
i\hbar &\frac{\partial}{\partial t}\tilde{u}(r,t)= -\frac{\hbar^2}{2 m}\frac{\partial^2}{\partial r^2}\bigg(\tilde{u}(r,t)\bigg)\CR
&+ \bigg(\frac{U_0}{r^2} |\tilde{u}(r,t) +r\sqrt{\rho}e^{-i\mu t/\hbar}|^2 + \sub{V}{Ryd,S,n}(r)\bigg)\tilde{u}(r,t)\CR
&+ \bigg( \frac{U_0}{r^2} |\tilde{u}(r,t) + r\sqrt{\rho}e^{-i\mu t/\hbar}|^2\CR
& + \sub{V}{Ryd,S,n}(r)\bigg)r\sqrt{\rho}e^{-i\mu t/\hbar}
-r\mu\sqrt{\rho}e^{-i\mu t/\hbar},
\end{eqnarray}
where we have used that the actual 3D density is $\rho(r,t) = |\frac{u(r,t)}{r}|^2  = |\frac{\tilde{u}(r,t)+r\sqrt{\rho}e^{-i\mu t/\hbar}}{r}|^2$.

Working with an asymptotically vanishing $\tilde{u}(r,t)$ was required for the final step, which is to anti-symmetrically expand $\tilde{u}(r,t)$ to negative $r$, such that $\tilde{u}(-r)=-\tilde{u}(r)$. This enforces $\frac{\partial^2}{\partial r^2}|_{r=0}=0$, thus preventing any cross-talk between the physical positive $r>0$ and the unphysical negative range. 

We verified the above transformations by a direct comparison of its results with complete 3D simulations.

\section{Truncated Wigner method}
\label{app_TWA}
%
The truncated Wigner method allows to investigate the dynamics of quantum depletion or thermal fluctuation, as long as these are small corrections to a strong mean-field. After the method's introduction to BEC~\cite{steel:wigner,Sinatra2001,castin:validity} the TWA in that context is described in many articles including the review \cite{blair:review}. The central ingredient of the method is adding random noise to the initial state of the GPE \bref{impurityGPE}. We thus use the initial \emph{stochastic field} 
\begin{eqnarray}
\alpha(\bv{R},0)&=\phi_0 + \sum_k [\eta_k u_k(\bv{R}) - \eta^*_k v^*_k(\bv{R}) ]/\sqrt{2}.
\label{twainistate}
\end{eqnarray}
with random complex Gaussian noises $\eta_k$ fulfilling $\overline{\eta_k\eta_l}=0$ and  $\overline{\eta_k\eta^*_l}=\delta_{nl}$, where $\overline{\dots}\mbox{ }$ is a stochastic average. $u_k(\bv{R})$ and $v_k(\bv{R})$ are the usual (3D) Bogoliubov modes in a homogeneous BEC with homogenous density $\rho=|\phi_0|^2$ \cite{book:pethik}.

A different symbol $\alpha(\bv{R})$ has been chosen for the stochastic field compared to the mean field $\phi(\bv{R})$, to emphasise the difference in physical interpretation due to the presence of noise: The stochastic field now allows the approximate extraction of quantum correlations using the prescription
\begin{eqnarray}
\frac{1}{2} \big(\expec{\hat{\Psi}^\dagger(\mathbf{R}')\hat{\Psi}(\mathbf{R})} + \expec{\hat{\Psi}(\mathbf{R}) \hat{\Psi}^\dagger(\mathbf{R}')} \big) \rightarrow  \overline{\alpha^*(\mathbf{R}') \alpha(\mathbf{R})},
\label{averages}
\end{eqnarray}
in which spatial correlations of the stochastic fields provide information on \emph{symmetrically ordered} quantum expectation values. In \bref{averages} $\hat{\Psi}(\mathbf{R})$ is the atomic field operator that destroys an atom at location $\mathbf{R}$ \cite{book:pethik}.

Using restricted basis commutators $\delta_c$ \cite{norrie:long,norrie:thesis}, we can then extract the total atom density 
\begin{eqnarray}
\sub{\rho}{t}(\mathbf{R})&=\overline{|\alpha(\mathbf{R})|^2}-\frac{\delta_c}{2},
\label{ntot}
\end{eqnarray}
condensate density $\sub{\rho}{c}(\mathbf{R})=\left|\overline{\alpha(\mathbf{R})} \right|^2$ and from these both the uncondensed density
$\sub{\rho}{u}(\mathbf{R})=\sub{\rho}{t}(\mathbf{R}) - \sub{\rho}{c}(\mathbf{R})$, see also \cite{wuester:nova2,wuester:kerr,wuester:collsoll}.
Uncondensed atom numbers as a measure of non-equilibrium ``heating" referred to in the main article are finally $\sub{N}{u} = \int d^3 \mathbf{R}\: \sub{\rho}{u}(\mathbf{R})$. 
\vspace{1cm}

%
\end{document}